# Inverse Microparticle Design for Enhanced Optical Trapping and Detection Efficiency in All Six Degrees of Freedom

Moosung Lee[1*], Benjamin A. Stickler[2], Thomas Pertsch[3,4], and Sungkun Hong[1*]

[1] *Institute for Functional Matter and Quantum Technologies and Center for Integrated Quantum Science and Technology, University of Stuttgart, 70569 Stuttgart, Germany;*

[2] *Institute for Complex Quantum Systems and Center for Integrated Quantum Science and Technology, Ulm University, 89069 Ulm, Germany;*

[3] *Institute of Applied Physics, Abbe Center of Photonics, Friedrich Schiller University Jena, 07745 Jena, Germany;*

[4] *Fraunhofer Institute for Applied Optics and Precision Engineering, 07745 Jena, Germany*

* *Corresponding authors:* M.L (moosung.lee@fmq.uni-stuttgart.de), S. H (sungkun.hong@fmq.uni-stuttgart.de)

**ABSTRACT**

Achieving quantum-limited motional control of optically trapped particles beyond the sub-micrometer scale is an outstanding problem in levitated optomechanics. A key obstacle is solving the light scattering problem and identifying particle geometries that allow stable trapping and efficient motional detection of their center of mass and rotational motion in three dimensions. Here, we present a computational framework that combines an efficient electromagnetic scattering solver with the adjoint method to inversely design printable microparticles tailored for levitated optomechanics. Our method allows identifying optimized geometries, characterized by enhanced optical trapping and detection efficiencies compared to conventional microspheres. This improves the feasibility of quantum-limited motional control of all translational and rotational degrees of freedom in a standard standing-wave optical trap.

## INTRODUCTION

Exploring the dynamics of optically trapped objects in vacuum has led to diverse applications in precision sensing and nonequilibrium physics[1,2]. A key frontier in this field is attaining quantum-limited motional control of levitated particles. Following early efforts to control and stabilize the translational motion of nanospheres at millikelvin temperatures[3–5], the field has recently achieved seminal breakthroughs, including cooling of a dielectric nanoparticle's motion to its quantum ground state via a cavity-assisted[6] or measurement-based scheme[7,8]. Recent advancements have further extended quantum control to the libration modes of anisotropic nanoparticles[9–12], even reaching high-purity ground states[13].

A crucial yet challenging next step in levitated optomechanics is extending quantum control to larger particles beyond sub-micrometer scales. Pushing this size limit is essential for macroscopic quantum experiments, such as matter-wave interferometry[14] and wavefunction collapse tests[15]. A prerequisite for this achievement is optimizing trapping[13] and motional detection efficiencies[16]. This challenge consists of two problems: (i) the forward problem of accurately computing the trapping and detection efficiencies for all observable translational and librational modes, and (ii) the inverse problem of configuring potential landscapes and particle structures to enhance these efficiencies. Solving both requires three-dimensional (3D) multiple scattering computations, which remain highly elusive beyond the Rayleigh regime[17,18] .

Recently, trapping particles beyond the Rayleigh regime has been numerically investigated in various contexts. For instance, Mie theory has been employed to analyze motional detection efficiencies[19,20] for a microsphere and its full 3D translational trap stability with varying sizes[21]. Others used boundary-based scattering solvers to analyze the 3D translational detection efficiency of cylinders and polygonal plates[22]. Such methods have also been used to identify particle structures with improved translational trap stiffness in 2D[23,24]. However, in practical applications, optimization must ensure simultaneous stable trapping and efficient motional detection across all six degrees of freedom in 3D. The six fundamental modes include three translational modes corresponding to the movement of a particle's center of mass and three libration modes described by rotation angles about principal axes. Addressing this inverse design problem requires general 3D electromagnetic scattering solvers, such as the finite-difference time-domain method[25], which remains practically demanding owing to its high computational costs.

Here, we present an efficient optimization algorithm for the inverse design of microparticle geometries tailored for quantum-limited levitated optomechanics experiments. Our computational framework integrates the modified Born series[26–28] for forward scattering calculations with the adjoint method[29–31] for inverse optimization. By incorporating shape constraints, we generate simplified 3D microstructures that are compatible with lithographic fabrication while ensuring 3D trap stiffness and detection efficiencies suitable for quantum-limited motional control in all translational and rotational degrees of freedom.

## Results

### Configuration of the optical trap and detection

Our inverse design framework considers a monolithic particle levitated within an optical trap operating at a total power of 250 mW and a wavelength of $\lambda$ = 1550 nm [Fig. 1(a)]. The optical trap is assumed to be a standing-wave trap formed by two counter-propagating, $x$-polarized Gaussian beams. This configuration eliminates the influence of scattering forces[32] and the torque induced by circular polarization[33]. Here, $\mathbf{E}_{\text{in}}(\mathbf{r}) = [\mathbf{E}_0(\mathbf{r}) + \mathbf{E}_0(-\mathbf{r})] / \sqrt{2}$ represents the complex field amplitude of an optical trap, where $\mathbf{E}_0(\mathbf{r})$ is the 3D electric field of a Gaussian beam propagating along the $z$-axis generated by a single objective lens with a numerical aperture (NA) of 0.8. We calculated $\mathbf{E}_0(\mathbf{r})$ using the vectorized angular spectrum method[18]. The design region is confined to a cylindrical volume aligned with the $y$-axis, with its diameter and height ranging from 2.2 to 2.8 μm. The diameter is chosen to include the three brightest intensity lobes in the XZ plane. The cylindrical shape of the design volume is selected to facilitate the implementation of certain geometric constraints, such as mirror symmetry and a restriction of the particle's shape to quasi-2D structures. The optical detection system is assumed to collect scattered light using the two objective lenses, which are also used for trapping.

## Principles of structure optimization

A key computational challenge in our inverse particle design is the efficient evaluation of the 3D total electric field. We address this challenge by employing a modified Born series and the adjoint method [Fig. 1(b)]. The modified Born series has been shown to outperform the finite-difference time-domain method in efficiently solving both forward[26,27] and inverse[28] scattering problems. We choose to combine this scattering solver with the adjoint method as an effective and efficient gradient-based inverse optimization algorithm[30,31].

The shape of a levitated particle in the simulation space is represented as a 3D binary connected map, $S(\mathbf{r})$, where its value is 1 inside the particle and 0 elsewhere. Accordingly, the relative permittivity map within the simulation domain is defined as $\varepsilon(\mathbf{r}) = 1 + \chi_e S(\mathbf{r})$, where $\chi_e$ denotes the bulk electric susceptibility of the designed particle. Solving the inhomogeneous wave equation for this permittivity distribution yields the 3D total electric field, $\mathbf{E}(\mathbf{r})$[27]:

$$\nabla \times [\nabla \times \mathbf{E}(\mathbf{r})] - k^2 \varepsilon(\mathbf{r})\mathbf{E}(\mathbf{r}) = \mathbf{0}, \tag{1}$$

where $k = 2\pi/\lambda$ is the wavenumber. The resulting total electric field is utilized to calculate the particle's translational and librational angular trap frequencies ($\Omega_j$, where $j$ denotes an index that labels all six degrees of freedom), and the corresponding detection efficiencies ($\eta_j$), and ultimately to determine the objective functional defined as a linear combination of these quantities along with a regulation term for the particle mass ($m$), centered around a target mass $m_0$ (see Supplementary Notes 1, 2 in Supporting Information (SI) for more details):

$$\mathcal{L}[\boldsymbol{\psi}(\mathbf{r}), S(\mathbf{r})] = \sum_j \{\Omega_j^2[\boldsymbol{\psi}(\mathbf{r}), S(\mathbf{r})] + \tau_\eta \eta_j[\boldsymbol{\psi}(\mathbf{r})]\} - \tau_m \{m[S(\mathbf{r})] - m_0\}^2 \tag{2}$$

Here, $\boldsymbol{\psi}(\mathbf{r}) = [\mathbf{E}(\mathbf{r}), \partial_j' \mathbf{E}(\mathbf{r}), \mathrm{c.c.}]$ represents the state variables of the fields dependent on the $S(\mathbf{r})$ and $\partial_j' \mathbf{E}(\mathbf{r})$ is the derivative of the total electric field with respect to an infinitesimal displacement of the permittivity distribution along the $j$-th motional degree of freedom. The regularization weights $\tau_\eta$ and $\tau_m$ control the contributions of $\eta_j$ and the mass regularization term respectively.

We now implement an iterative gradient approach based on the adjoint method [30,31]. To enable gradient update, we allow the values of $S(\mathbf{r})$ to vary continuously from 0 to 1 during optimization. The initial condition is set using a small ellipsoid at the center with surrounding regions populated by random Gaussian noise, following common random initialization strategies in machine learning [34]. The shape function and the corresponding relative permittivity map are then iteratively updated using the gradient of the objective functional, as derived in Supplementary Note 3 in SI.

Here, to steer $S(\mathbf{r})$ to converge close to a fully connected 3D binary structure, we impose several post processing steps at every iteration. Specifically, we use Gaussian blurring to enforce a minimum feature size and apply adaptive binarization to facilitate faster convergence to a monolithic binary structure[30] (see Supplementary Notes 4 and 5 in SI). In addition, mirror symmetry is imposed to ensure that the axes of the optical trap are aligned with the particle's principal axes and, thus, those of translational and librational normal modes. After convergence, the final 3D shape is obtained by binarizing $S(\mathbf{r})$ with a threshold of 0.5 and removing any unconnected pixelated artefacts outside the trap center arising from thresholding [Fig. 1(c)].

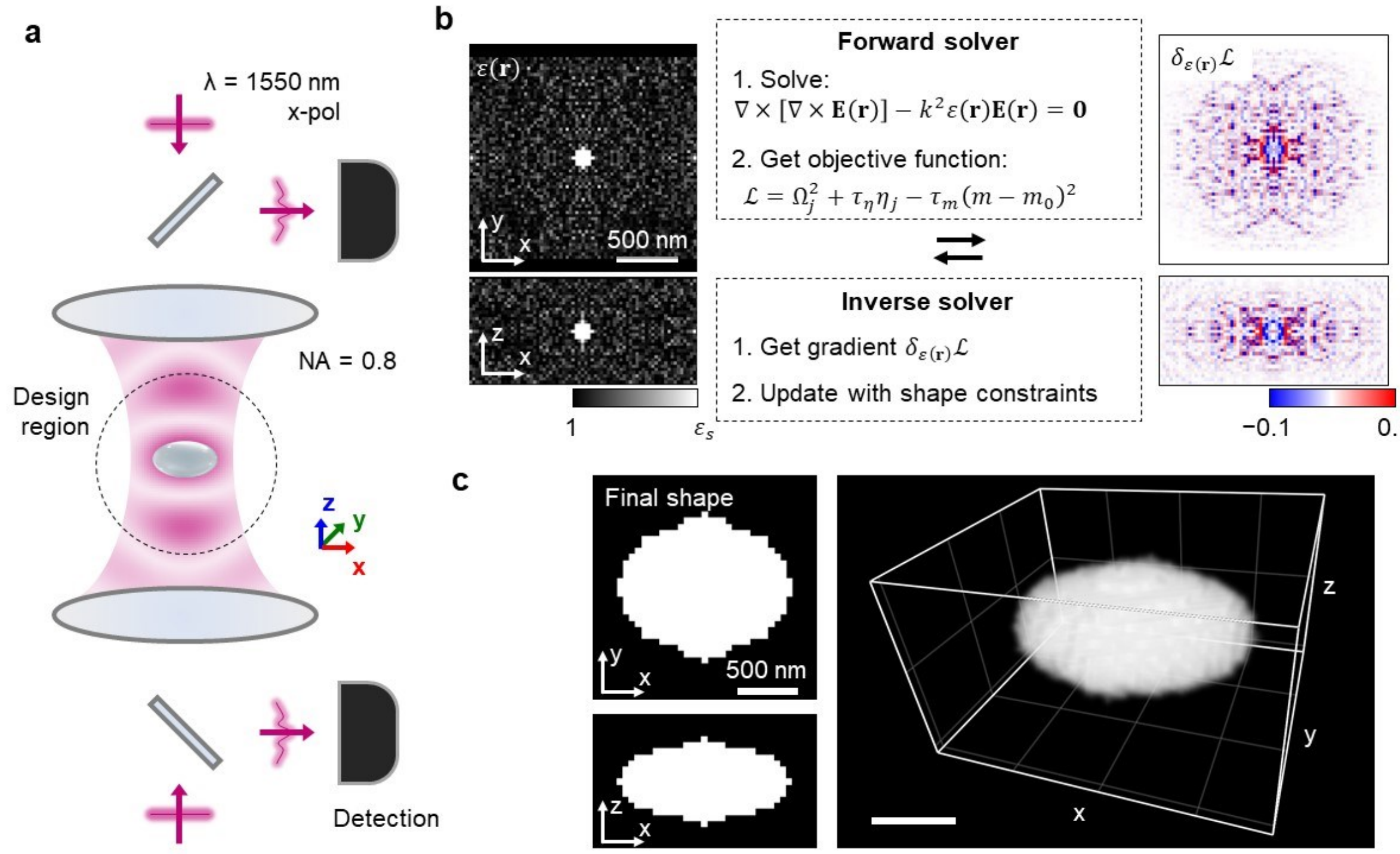

**Figure 1 | Overview of the inverse design framework. (a)** Optical configuration. A simulation setup featuring a monochromatic optical trap with a total power of 250 mW and a wavelength of 1550 nm. The standing-wave optical trap is formed by two counter-propagating, *x*-polarized Gaussian beams, with scattered light collected from both forward and backward propagation directions using two objectives. The numerical aperture (NA) is set to 0.8 in both directions. **(b)** Gradient-based optimization. The process begins with random initialization by embedding a dielectric ellipsoid at the center in surround regions populated by random Gaussian noise. The relative permittivity is iteratively updated through a sequential forward and inverse process. In the forward step, the total electric field, $\mathbf{E}(\mathbf{r})$, is computed by solving the 3D inhomogeneous wave equation, yielding an objective functional defined as a combination of squared angular trap frequencies ($\Omega_j$), detection efficiencies ($\eta_j$), and a mass regulation term centered around a target mass $m_0$. In the inverse step, the adjoint method calculates the gradient of the objective functional while imposing shape constraints, including mirror symmetry, smoothing, adaptive binarization, and thickness fixation. **(c)** A representative 3D silica particle structure produced by the optimization process.

**Benchmarking performance with 3D silica particle design**

To evaluate the effectiveness of our optimization method, we design 3D binary silica ($SiO_2$) microstructures under two optimization conditions: (i) optimizing only $\Omega_j$, and (ii) optimizing both $\Omega_j$ and $\eta_j$ [Fig. 2]. By setting the targeted mass $m_0$ to 4.7 fg, the target volume is constrained to approximately 1.6 $\mu m^3$ [Fig. 2(a)]. Both results exhibit complex 3D morphologies with substantial overlaps with the three brightest sidelobes.

To quantitatively assess the performance enhancement, we compare the optical forces and torques exerted on the optimized particles with those on a microsphere of the same volume [Fig. 2(b)]. Due to mirror symmetry, the directions of the optical forces acting on the particle at the trap focus are aligned with the translational displacements with respect to the lab frame axes ($x$, $y$, $z$). This symmetry also constrains the translational modes to align with the particle's principal axes and the directions of the resulting optical torques with respect to rotations about these axes—namely, ($\gamma$, $\beta$, $\alpha$) orientations. Our simulations reveal that the $SiO_2$ microsphere exhibits weak axial trap stiffness, while it experiences restoring forces with trap frequencies exceeding 100 kHz along the radial directions. We note that optical torques do not appear for microspheres due to their spherical symmetry. In contrast, both of the optimized structures show strong stiffness across all translational and rotational degrees of freedom. Corresponding frequencies of the particle's translational and librational motions range from 90 to 280 kHz, along with detection efficiencies exceeding 46% across all motional degrees of freedom.

To further investigate the size dependence of the optimization performance, we analyze optimized structures across different volumes [Fig. 3]. Compared to spherical particles of the same volume, the optimized structures generally exhibit higher translational trap frequencies and detection efficiencies, particularly along the $z$-axis. Notably, the optimized structures maintain stable trapping even in size regimes where trapping spherical particles becomes unstable at the focus due to negative trap stiffness along the $z$-axis. Regarding optimization, explicitly optimizing $\eta$ has only a marginal effect, despite its significantly higher computational cost due to the far-field computation process. Based on these analyses, we opt to optimize only the trap frequencies while ensuring that motional detection efficiencies exceeded 20%, a threshold sufficient for ground-state cooling via measurement-based control[16].

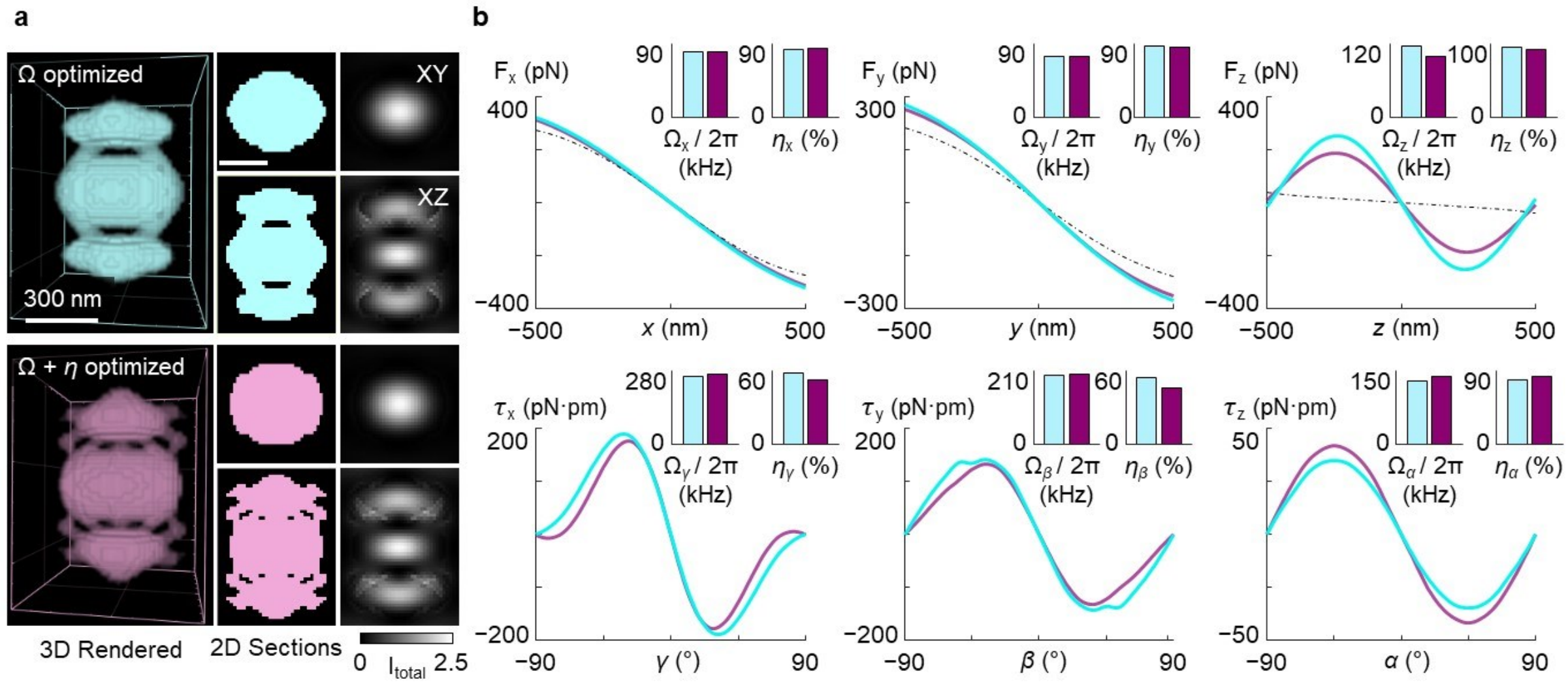

**Figure 2 | Optimized 3D $SiO_2$ microstructures and their optical trapping and detection characteristics. (a)** Optimized $SiO_2$ structures obtained through two distinct strategies: optimizing only the angular trap frequencies (top; cyan) or both the angular trap frequencies and detection efficiencies (bottom; magenta). Left: 3D rendered tomograms of the optimized structures. Middle and right: 2D cross-sectional views (XY and XZ planes) of the structure (middle) and the corresponding total electric field intensity, $I_{total}$ (right), across the particle's center. **(b)** Optical forces (top panels) and torques (bottom panels) as functions of translational ($x$, $y$, $z$) and rotational ($\gamma$, $\beta$, $\alpha$) displacements with respect to the three lab frame axes. Due to mirror symmetry, particles' principal axes coincide with the lab frame axes. The dotted grey lines represent the optical force profile for a spherical particle of equivalent volume (~1.6 μm$^3$) for comparison. We note that optical torques for a spherical particle do not exist and are thus not shown in the bottom panels. Insets: trap frequencies ($\Omega_j/2\pi$) and motional detection efficiencies ($\eta_j$) at the trap focus for each optimized structure.

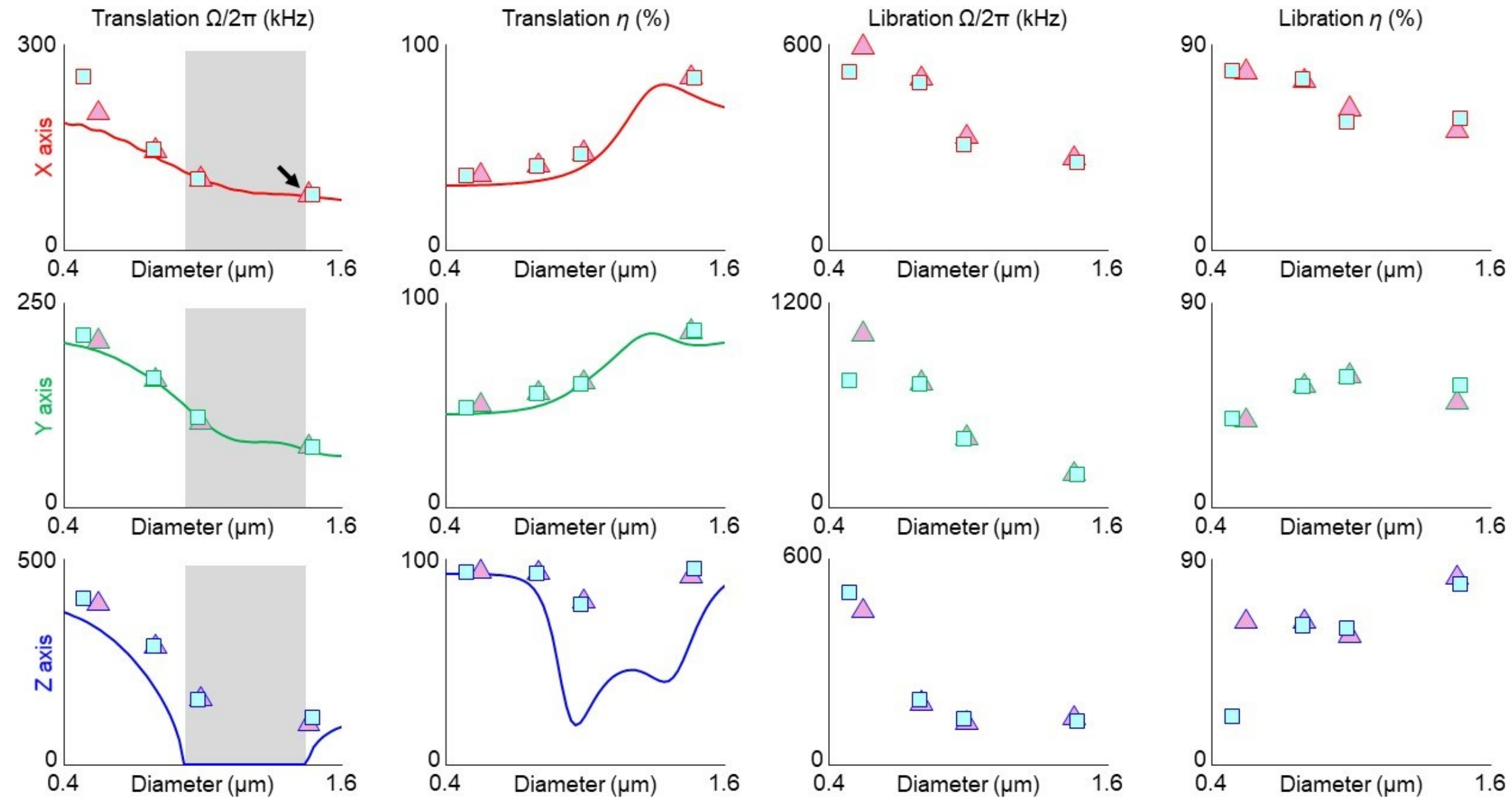

**Figure 3 | Size dependence of 3D optimization on optical trapping and detection efficiency for $SiO_2$ 3D microstructures.** Markers represent parameters for optimized structures (cyan square: Ω-optimized, and magenta triangle: $\Omega + \eta$-optimized). The black arrow marks the data points used in Fig. 2. First and second columns: Lines indicate translational trap frequency (column 1) and detection efficiency (column 2) of a $SiO_2$ sphere as functions of diameter. Shaded areas indicate size ranges where spherical particles cannot be stably trapped. Third and fourth columns: librational trap frequency (column 3) and detection efficiency (column 4) for the optimized structures. Each row corresponds to a different axis ($x$, $y$, or $z$).

## Shape optimization of 3D extruded structure

A major challenge in experimentally applying our method lies in whether our optimization algorithm can generate structures that are readily fabricable in a scalable manner. Conventional micro- or nano-fabrication technologies based on lithography are largely restricted to producing quasi-2D extruded structures[35]. This necessitates the implementation of appropriate shape constraint strategies in the optimization process. To address this, we introduce thickness fixation constraints and optimize the base geometry to generate candidate 3D extruded structures (Figs. 4A and B; see also Supplementary Notes 4 and 5 in SI). The orientation of the particle's base is constrained to remain parallel to the XZ plane.

The optimized results yield quasi-2D $SiO_2$ particles with higher trap frequencies and motion detection efficiencies than spheres of the same volume [Fig. 4(c)]. Similar to the result in Fig. 3, the extruded structures remain stably trapped in size regimes where trapping spherical particles is unstable along the *z*-axis. Notably, a comparison with the results in Fig. 3 shows that imposing this additional extrusion geometry constraint does not significantly affect the optimization performance, validating the effectiveness of our shape constraint strategy.

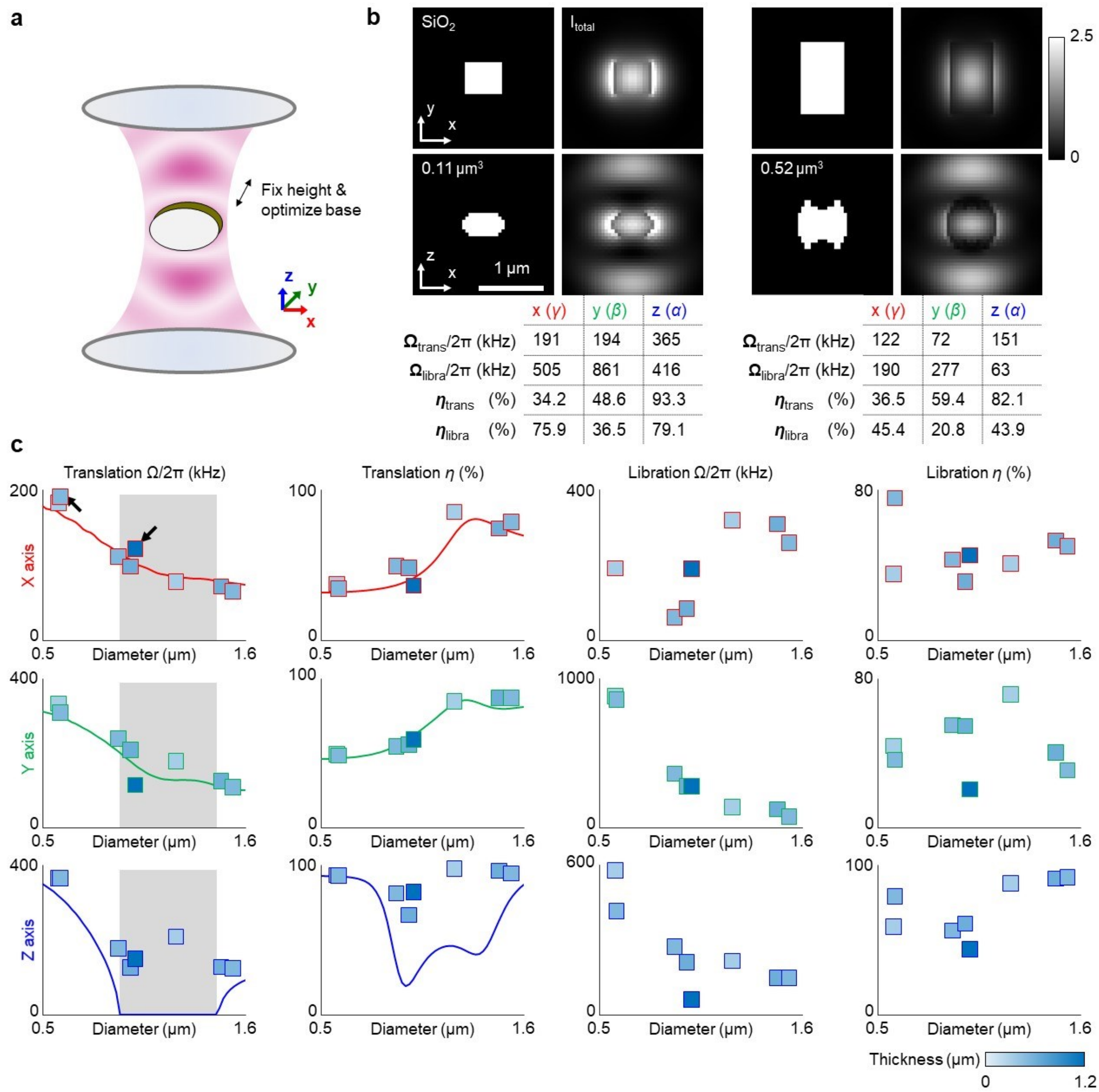

**Figure 4 | Optimization results of extruded $SiO_2$ structures. (a)** Thickness constraints in the inverse design. A particle of a fixed height is assumed to be aligned along the y-axis, and its base geometry is optimized to enhance trapping frequencies across all translational and rotational modes. **(b)** 2D cross-sectional views (XY and XZ planes) of the structure (left) and the corresponding total electric field intensity, $I_{total}$ (right), across the particle's center. The estimated trap frequencies and detection efficiencies are given in tables. **(c)** Optimization parameters as functions of sphere diameter of equivalent volume. Marker colors indicate the thickness of each optimized structure. The black arrows mark the data points used in (b). Each column shows translational trap frequency (1st column), its corresponding detection efficiency (2nd column), the librational trap frequency (3rd column), and the corresponding detection efficiency (4th column). Shaded areas denote the size regime where spheres cannot be trapped. Lines indicate the translational trap frequency (column 1) and detection efficiency (column 2) of a sphere as functions of diameter. Each row corresponds to a different axis (*x*, *y*, or *z*).

## Shape optimization of printable silicon particles

While we have investigated silica particles, Silicon (Si) is another promising material for levitated optomechanics, offering compatibility with nanolithography techniques[36] and a susceptibility nearly ten times higher ($\chi_{e,\,Si} = 11.1$) than that of $SiO_2$ ($\chi_{e,\,SiO2} = 1.07$). To explore its potential, we extend our inverse design framework to optimize Si-based extruded structures [Fig. 5(a)]. The high susceptibility induces pronounced resonance effects, enabling the optimized particles to function as levitated optical resonators. These resonances lead to a significant increase in trapping frequencies, particularly those associated with librational motions, even for a microparticle with a volume of 0.92 $\mu m^3$.

To assess the effect of structure optimization, we plot the trapping performance as a function of sphere diameter of equivalent volume [Fig. 5(b)]. Compared to $SiO_2$, spherical Si particles experience unstable trapping conditions at the trap focus more frequently across a selected range for sizes, underscoring the necessity of precise selection of particle size and geometry in a given high-NA trap condition[33,21]. In contrast, our optimization algorithm allows us to find the shapes of particles exhibiting stable trapping over a wide range of sizes. Furthermore, the Si particles yield significantly high librational trap frequencies, exceeding 1 MHz. These enhancements are expected to be highly advantageous for achieving high-purity motional ground states via sideband-resolved cooling schemes[13].

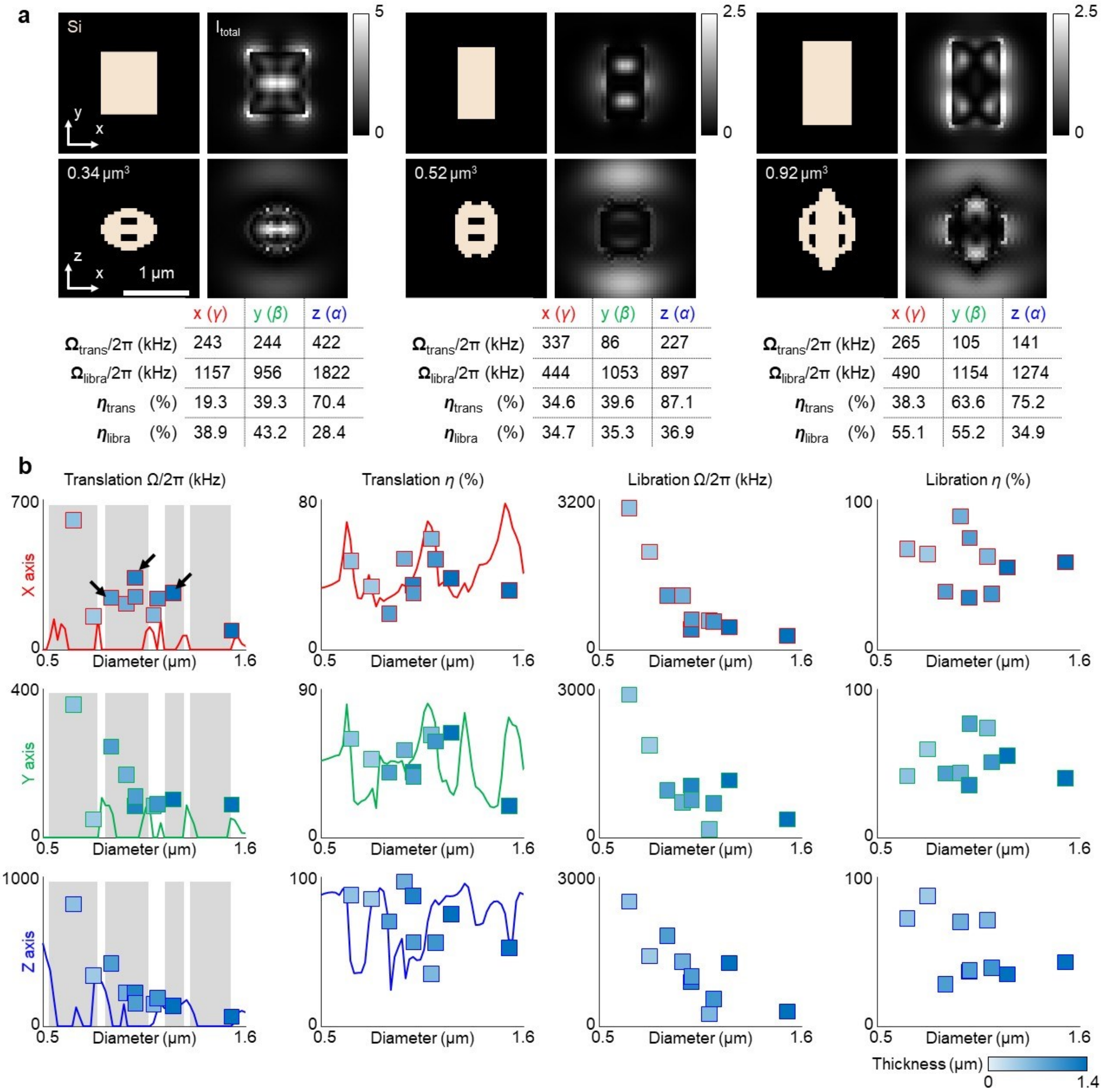

**Figure 5 | Optimization results of extruded Si structures. (a)** Optimization results for particle volume of 0.34 µm³ (1st column), 0.52 µm³ (2nd column), and 0.92 µm³ (3rd column). Shown are 2D cross-sectional views (XY and XZ planes) of the structure (left) and the corresponding total electric field intensity, $I_{total}$ (right), across the particle's center. The estimated trap frequencies and detection efficiencies are given in tables. **(b)** Optimization parameters as functions of particle diameter of a sphere of equivalent volume. Marker colors indicate the thickness of each optimized structure. The black arrows mark the data points used in (a). Each column shows translational trap frequency (1st column), its corresponding detection efficiency (2nd column), the librational trap frequency (3rd column), and the corresponding detection efficiency (4th column). Shaded areas denote the size regime where spheres cannot be trapped. Lines indicate the translational trap frequency (column 1) and detection efficiency (column 2) of a sphere as functions of diameter. Each row corresponds to a different axis (*x*, *y*, or *z*).

## Discussion

A major obstacle in quantum levitated optomechanics beyond the Rayleigh regime is engineering light-matter interactions to achieve effective optical trapping and motion detection across as many motional degrees of freedom as possible[10,11,37]. This challenge largely stems from the difficulty of efficiently evaluating optical trapping conditions in the presence of multiple light scattering. Our work addresses this issue by integrating advanced techniques for solving forward and inverse electromagnetic scattering problems into levitated optomechanics[28].

Our numerical analyses offer valuable insights that complement earlier studies on levitated optomechanics with particles beyond Rayleigh nanospheres. In particular, our forward simulations yield results consistent with Mie theory regarding the translational trap frequencies[21] and motional detections[19,20] of microspheres in standing-wave optical traps. Moreover, our analysis reveals that the large susceptibility contrast of Si particles can hinder stable trapping at the trap focus with a simple spherical particle geometry as predicted previously[21]. Our study show that this tap stability can be recovered by shape optimization for any given volume.

We anticipate that our optimization approach will offer distinct advantages when assessing non-spherical particles for quantum levitated optomechanics. Specifically, our method unlocks a size regime in which simple spherical particles cannot be stably trapped. Moreover, the tailored structures obtained from our optimization method exhibit high libration frequencies—several times greater than those observed in nanosphere clusters[11,13,38] or unoptimized nanorods[33] used in current state-of-the-art experiments. This enhancement in stiffness offers clear benefits, including more efficient trapping with reduced power requirements and, therefore, lower photothermally induced trap instabilities[39]. Importantly, utilizing larger masses with a greater likelihood of achieving quantum-limited control can facilitate fundamental tests of quantum principles[14] and quantum-enabled sensing[40,41] with masses exceeding current records[42].

We note that the previously mentioned trap instability at the focus does not necessarily imply that the particle cannot be trapped elsewhere within the trapping field. For instance, Lepeshov et al.[21]

demonstrated that the onset of instability at the intensity maximum (i.e., the trap focus) can be accompanied by the emergence of a stable trapping condition near the intensity minima. While the present study focuses on optimizing the particle shape for trapping near the intensity maximum, we anticipate that the same method can be readily extended to design shapes suitable for trapping at intensity minima as well.

Looking ahead, future work will seek to enhance our approach across diverse domains. On the photonics front, promising directions include optimizing the 3D trap landscape by designing metasurfaces[43] or implementing wavefront shaping techniques[44–46]. Additional avenues for improvement involve refining shape constraints[47] and accelerating gradient computations[48] to further boost optimization performance. With our algorithm implemented on a graphics processing unit, we foresee the use of computational clusters to extend these optimization problems to even larger mass scales[49].

**Acknowledgements**

This research was supported by the Carl-Zeiss-Stiftung (CZS) via the CZS Center for Quantum Photonics (QPhoton) and by Ministry of Science, Research and Arts of Baden-Württemberg, Germany. M.L. acknowledges the support funded by Alexander von Humboldt Foundation. B.A.S. acknowledges funding by the German Research Foundation (DFG; 510794108).

**Conflict of interest**

The authors declare no conflict of interest.

**Data Availability**

The presented optimisation results are available in Github: https://github.com/moosunglee/particle_opt.

**Supporting Information**

The Supporting Information is available free of charge: theoretical derivations of electromagnetic objective functional, its discrete form and gradient, shape constraints, and pseudo-code.

# Inverse Microparticle Design for Enhanced Optical Trapping and Detection Efficiency in All Six Degrees of Freedom: Supporting Information

## Supplementary Note 1. Electromagnetic objective functional

This section revisits the electromagnetic equations governing trap stiffness and detection efficiencies. To compute these parameters, it is necessary to solve the inhomogeneous wave equation (Eq. (1) in the main text). Given the shape function $S(\mathbf{r})$ defined in the main text and the incident monochromatic trapping field $\mathbf{E}_{in}(\mathbf{r})$, the total electric field satisfies the following integral equation[1,2]:

$$\mathbf{E}(\mathbf{r}) = \mathbf{E}_{in}(\mathbf{r}) + k^2\chi_e \int \overleftrightarrow{\mathbf{G}}(\mathbf{r}-\mathbf{r}')\cdot S(\mathbf{r}')\mathbf{E}(\mathbf{r}')d^3\mathbf{r}', \qquad \text{(S1)}$$

here $S(\mathbf{r})$ ensures the integral covers only the particle volume. Here, $\overleftrightarrow{\mathbf{G}}(\mathbf{r}) = \left[\mathbf{I}_3 + \frac{1}{k^2}\nabla\nabla^T\right]\frac{e^{ikr}}{4\pi r}$ is the dyadic Green's tensor, $r = |\mathbf{r}|$, and $\mathbf{I}_n$ is the $n \times n$ identity matrix.

From the 3D total electric field, the time-averaged radiation force ($F_j$) and torque ($T_j$) along the $j$-th ($x$, $y$, or $z$) axis at the origin are given by the following functionals[3,4]:

$$F_j[\mathbf{E}(\mathbf{r}),\mathbf{E}^*(\mathbf{r}),S(\mathbf{r})] = -\frac{\varepsilon_0\chi_e}{4}\int\left[\partial_j S(\mathbf{r})\right]|\mathbf{E}(\mathbf{r})|^2 d^3\mathbf{r}, \qquad \text{(S2)}$$

$$T_j[\mathbf{E}(\mathbf{r}),\mathbf{E}^*(\mathbf{r}),S(\mathbf{r})] = -\frac{\varepsilon_0\chi_e}{4}\int[(\mathbf{r}-\mathbf{r}_0)\times\nabla S(\mathbf{r})]_j|\mathbf{E}(\mathbf{r})|^2 d^3\mathbf{r}, \qquad \text{(S3)}$$

where $\mathbf{r}_0$ is the center-of-mass position. Our constraint enforces the particle geometry to be symmetric with respect to the XY, YZ, and XZ planes, leading $\mathbf{r}_0$ to be $\mathbf{0}$. This symmetry also ensures that the particle's principal axes are aligned with the fixed spatial $x$, $y$, and $z$ axes. Consequently, the orientations of the fundamental translational and librational modes are also aligned with the lab frame's axes when it is trapped at the focus of a linearly polarized standing-wave optical trap. The trapping frequencies of these modes are then derived from the trap stiffness along the $j$-th translational ($\kappa_j$) and rotational axes ($J_j$), which are proportional to the derivatives of $F_j$ and $T_j$, with respect to the infinitesimal translational and rotational displacement of the particle about the $j$-th axis respectively:

$$\kappa_{j\in(x,y,z)} = -\partial_j' F_j = \lim_{\delta x\to 0}\frac{F_j[S(\mathbf{r})]-F_j[S(\mathbf{r}-\delta x\hat{\jmath})]}{\delta x},, \tag{S4}$$

$$J_{j\in(x,y,z)} = -\partial_j' T_j = \lim_{\delta\theta\to 0}\frac{T_j[S(\mathbf{r})]-T_j[S(\boldsymbol{R}(-\delta\theta\hat{\jmath})\cdot\boldsymbol{r})]}{\delta\theta}, \tag{S5}$$

where $\hat{\jmath}$ is a unit vector parallel to the $j$-th axis and $\mathbf{R}(\delta\boldsymbol{\theta})$ represents the rotation operation about the axis parallel to $\delta\boldsymbol{\theta}$ by an angle $|\delta\boldsymbol{\theta}|$. The angular trap frequencies for the translational and librational modes are defined as $\Omega_{j,\mathrm{trans}} = \sqrt{\kappa_j/m}$ and $\Omega_{j,\mathrm{rot}} = \sqrt{J_j/I_j}$, respectively, where $m$ is the particle's mass and $I_j$ is the moment of inertia about $j$-th axis, respectively. We estimate $m$ and $I_j$ under the assumption that the mass density is proportional to the particle's shape function:

$$m = \rho_0 \int S(\mathbf{r}) d^3\mathbf{r}, \tag{S6}$$

$$J_{j\in(x,y,z)} = -\partial_j' T_j = \lim_{\delta\theta\to 0}\frac{T_j[S(\mathbf{r})]-T_j[S(\mathbf{R}(-\delta\theta\hat{\jmath})\cdot\mathbf{r})]}{\delta\theta}, \tag{S7}$$

where $\rho_0$ is the density-to-susceptibility ratio and $x_j$ is the displacement along the $j$-th axis.

The motion detection efficiency $\eta_j$ is determined based on the far-field intensity associated with the Fisher information about the particle's $j$-th motional degree of freedom, $\left|\partial_j'\mathbf{E}(\hat{\mathbf{k}})\right|^2$, where $\hat{\mathbf{k}} = (\hat{\mathbf{k}}_\perp, \hat{k}_z) = (\hat{k}_x, \hat{k}_y, \hat{k}_z)$ is the unit vector in the direction of information radiation[5,6]. $\partial_j'\mathbf{E}$ is defined as the derivative of the electric field with respect to the infinitesimal translational and rotational displacement of the particle about the $j$-th axis respectively:

$$\partial_{j,trans}'\mathbf{E} = \lim_{\delta x\to 0}\frac{\mathbf{E}[S(\mathbf{r}-\delta x\hat{\jmath})]-\mathbf{E}[S(\mathbf{r})]}{\delta x}, \tag{S8}$$

$$\partial_{j,rot}'\mathbf{E} = \lim_{\delta\theta\to 0}\frac{\mathbf{E}[S(\mathbf{R}(-\delta\theta\hat{\jmath})\cdot\mathbf{r})]-\mathbf{E}[S(\mathbf{r})]}{\delta\theta}. \tag{S9}$$

Under our simulation conditions, the detection efficiency is defined as the ratio of the information radiation collected by two objective lenses with a numerical aperture (NA) of 0.8 to the total information radiation power:

$$\eta_j\left[\partial_j'\mathbf{E}, \partial_j'\mathbf{E}^*\right] = \frac{\int_{(|\hat{\mathbf{k}}|=1)\cap(|\hat{\mathbf{k}}_\perp|\le \mathrm{NA})}\left|\partial_j'\mathbf{E}(\hat{\mathbf{k}})\right|^2 d^2\hat{\mathbf{k}}}{\int_{|\hat{\mathbf{k}}|=1}\left|\partial_j'\mathbf{E}(\hat{\mathbf{k}})\right|^2 d^2\hat{\mathbf{k}}}. \tag{S10}$$

In realistic measurement settings, a more sophisticated analysis would need to account for the mode-matching efficiency and the effect of specific measurement schemes, such as homodyne detection[7]. For simplicity, this work focuses solely on the information collection efficiency.

## Supplementary Note 2. Discretization of objective functions

In a simulation space consisting of $N$ voxels, the 3D shape function is represented as a discrete $N \times 1$ vector, $S(\mathbf{r}) \equiv \mathbf{s}$. The total electric field is also discretized as a $3N \times 1$ vector, $\mathbf{E}(\mathbf{r}) \equiv \mathbf{e} = \left[\mathbf{e}_x^T, \mathbf{e}_y^T, \mathbf{e}_z^T\right]^T$, where 3 arises from *x*-, *y*-, and *z*-components of the electric field. The inhomogeneous wave equation (Eq. (1) of the main text) is expressed using the following differential operator on the field vector:

$$\hat{\mathbf{h}}[\mathbf{s}]\mathbf{e} = \nabla \times [\nabla \times \mathbf{e}] - k^2[\mathbf{I}_3 \otimes \mathrm{diag}(1 + \chi_e \mathbf{s})]\mathbf{e} = \mathbf{0}, \tag{S11}$$

where $\otimes$ is the Kronecker product. Given the discrete incident electric field, $\mathbf{E}_{in}(\mathbf{r}) \equiv \mathbf{e}_{in}$, its solution satisfies the discrete form of the integral equation, Eq. (A1). The numerical solution can be efficiently computed using the modified Born series, which was implemented and validated in our previous study[8].

Our optimization parameters, $\Omega_j$, $\eta_j$, and $m$, can also be represented as discrete functionals of $\mathbf{e}$ and $\mathbf{s}$. According to Eq. (S6), the mass term $m$ depends only on the shape function $\mathbf{s}$, which is simply discretized as:

$$m[\mathbf{s}] = \rho_0 \delta x^3 \mathrm{tr}[\mathrm{diag}(\mathbf{s})], \tag{S12}$$

where $\delta x^3$ reflects the volume of each voxel. The moment of inertia along the $j$-axis [Eq. (S7)] is discretized as:

$$I_j[\mathbf{s}] = \rho_0 \delta x^3 \left(|\mathbf{x}|^2 - \mathbf{x}_j^2\right)^T \cdot \mathbf{s}, \tag{S13}$$

where $\mathbf{x}$ is a $N \times 3$ vector representing the 3D displacement of $N$ voxels from the origin, and $\mathbf{x}_j$ represents the displacement along the $j$-th axis.

To derive the discrete formula of the squared angular trap frequency, $\Omega_j^2$, we define the trap stiffness in the discrete space, $\kappa_j = \left\{F_j[\mathbf{s} \equiv S(\mathbf{r} + \delta x \hat{j})] - F_j[\mathbf{s} \equiv S(\mathbf{r} - \delta x \hat{j})]\right\}/(2\delta x)$ and, $J_j =$ $\{T_j[\mathbf{s} \equiv S(\mathbf{R}(\delta\theta\hat{j}) \cdot \mathbf{r})] - T_j[\mathbf{s} \equiv S(\mathbf{R}(-\delta\theta\hat{j}) \cdot \mathbf{r})]\}/(2\delta\theta)$ where $\delta x$ and $\delta\theta$ are arbitrary values for finite differentiation. Here, the radiation forces and torques in the discrete domain are expressed as the following functionals:

$$F_j[\mathbf{s} \equiv S(\mathbf{r} - \delta\mathbf{r})] = -\chi_e \mathbf{e}^\dagger [\mathbf{I}_3 \otimes \mathrm{diag}(\partial_j \mathbf{s})] \mathbf{e}\big|_{\mathbf{s} \equiv S(\mathbf{r}-\delta\mathbf{r})}, \tag{S14}$$

$$T_j[\mathbf{s} \equiv S(\mathbf{R}(-\delta\boldsymbol{\theta}) \cdot \mathbf{r})] = -\chi_e \mathbf{e}^\dagger \{\mathbf{I}_3 \otimes \mathrm{diag}[(\mathbf{x} \times \nabla\mathbf{s})_j]\} \mathbf{e}\big|_{\mathbf{s} \equiv S(\mathbf{R}(-\delta\boldsymbol{\theta}) \cdot \mathbf{r})}. \tag{S15}$$

Here, we assume that $\mathbf{e}$ is re-normalized to incorporate the laser power, the discrete integration coefficient, $\delta x^3$, and a constant factor, $\varepsilon_0/4$ from Eqs. (S2) and (S3).

Lastly, the motional detection efficiency $\eta_j$ is calculated in discrete space by numerically integrating the far-field intensity of the Fisher information radiation:

$$\eta_j[\partial_j' \mathbf{e}, \partial_j' \mathbf{e}^*] = \frac{(\partial_j' \mathbf{e})^\dagger \mathbf{P}_{det} (\partial_j' \mathbf{e})}{(\partial_j' \mathbf{e})^\dagger \mathbf{P}_{tot} (\partial_j' \mathbf{e})}. \tag{S16}$$

Here, $\mathbf{P}_{tot} = \mathbf{U}^\dagger \{\mathbf{I}_3 \otimes \mathrm{diag}[\delta(|\mathbf{q}| - k)]\} \mathbf{U}$ is a projection operator representing the surface integration over the whole solid angle. Specifically, $\mathbf{U}$ represents the unitary Fourier transform operation mapping the real-space coordinate $\mathbf{r} = (x, y, z)$, to the Fourier-space coordinate $\mathbf{q} = (\mathbf{q}_\perp, q_z) = (q_x, q_y, q_z)$. $\delta(|\mathbf{q}| - k)$ is a Kronecker-Delta that selectively filters the far-field component in the Fourier space. $\mathbf{P}_{det} = \mathbf{U}^\dagger \{\mathbf{I}_3 \otimes \mathrm{diag}[\delta(|\mathbf{q}| - k)\Theta(k \cdot \mathrm{NA} - |\mathbf{q}_\perp|)]\} \mathbf{U}$ represents the surface integration over the detection range, and $\Theta(x)$ is a Heaviside function.

## Supplementary Note 3. Gradient of discrete objective functionals

We derive the gradient of each parameter contribution from the objective functional defined in Eq. (2) of the main text in the discrete form.

### A. Gradient of mass regulator

The mass is defined solely as a function of the discrete shape vector $\mathbf{s}$. Accordingly, the gradient of the mass regularization term with respect to the $n$-th component, $s_n$, is given by the following relation:

$$\frac{\delta}{\delta s_n}[-(m-m_0)^2] = -2(m-m_0)\frac{\delta m}{\delta s_n} = -2\rho_0 \delta x^3 (m-m_0). \qquad \text{(S17)}$$

### B. Gradient of squared angular trap frequency

The derivative of $\Omega_j^2[\mathbf{e}, \mathbf{e}^*, \mathbf{s}]$ with respect to $s_n$ is expressed:

$$\frac{\delta \Omega_{j,trans}^2}{\delta s_n} = \frac{\delta}{\delta s_n}\left(\frac{\kappa_j}{m}\right) = -\frac{\Omega_{j,trans}^2}{m}\frac{\delta m}{\delta s_n} + \frac{1}{2m\delta x}\left[\left.\frac{\delta F_j}{\delta s_n}\right|_{s\equiv S(\boldsymbol{r}+\delta x\hat{\jmath})} - \left.\frac{\delta F_j}{\delta s_n}\right|_{s\equiv S(\boldsymbol{r}-\delta x\hat{\jmath})}\right], \qquad \text{(S18)}$$

$$\frac{\delta \Omega_{j,rot}^2}{\delta s_n} = \frac{\delta}{\delta s_n}\left(\frac{J_j}{I_j}\right) = -\frac{\Omega_{j,rot}^2}{I_j}\frac{\delta I_j}{\delta s_n} + \frac{1}{2I_j\delta\theta}\left[\left.\frac{\delta T_j}{\delta s_n}\right|_{\mathbf{s}\equiv S(\mathbf{R}(\delta\theta\hat{\jmath})\cdot\mathbf{r})} - \left.\frac{\delta T_j}{\delta s_n}\right|_{\mathbf{s}\equiv S(\mathbf{R}(-\delta\theta\hat{\jmath})\cdot\mathbf{r})}\right], \qquad \text{(S19)}$$

where the gradient of the mass term is $\rho_0\delta x^3$ for the translational mode and $\rho_0\delta x^3\left(|\mathbf{x}|^2 - \mathbf{x}_j^2\right)_n$ for the librational mode, respectively.

The functional derivatives of the radiation force and torque are derived using the chain rule:

$$\frac{\delta F_j}{\delta s_n} = \frac{\partial F_j}{\partial s_n} + \frac{\partial F_j}{\partial \boldsymbol{\psi}_\Omega}\cdot\frac{\delta \boldsymbol{\psi}_\Omega}{\delta s_n}, \quad \frac{\delta T_j}{\delta s_n} = \frac{\partial T_j}{\partial s_n} + \frac{\partial T_j}{\partial \boldsymbol{\psi}_\Omega}\cdot\frac{\delta \boldsymbol{\psi}_\Omega}{\delta s_n}, \qquad \text{(S20)}$$

where $\boldsymbol{\psi}_\Omega = \left[\mathbf{e}^T, \mathbf{e}^\dagger\right]^T \in \mathbb{C}^{6N\times 1}$ is the state variable of the fields, dependent upon $\mathbf{s}$. To efficiently compute the functional derivative $\delta\boldsymbol{\psi}_\Omega/\delta s_n$, we employ a Lagrangian functional, $\mathbf{H}_\Omega[\mathbf{e}, \mathbf{e}^*, \mathbf{s}] = \left[\left(\hat{\mathbf{h}}[\mathbf{s}]\mathbf{e}\right)^T, \text{c.c.}\right]^T = \mathbf{0}$, whose gradient satisfies:

$$\frac{\delta \mathbf{H}_\Omega}{\delta s_n} = \frac{\partial \mathbf{H}_\Omega}{\partial s_n} + \frac{\partial \mathbf{H}_\Omega}{\partial \boldsymbol{\psi}_\Omega}\cdot\frac{\delta \boldsymbol{\psi}_\Omega}{\delta s_n} = \mathbf{0}, \qquad \text{(S21)}$$

where $\partial\mathbf{H}_\Omega / \partial\boldsymbol{\psi}_\Omega$ is a 2 × 2 block matrix, defined according to the convention of matrix calculus[9]. The relation allows us to substitute the functional derivative in Eq. (S20) with partial gradients:

$$\frac{\delta F_j}{\delta s_n} = \frac{\partial F_j}{\partial s_n} - \underbrace{\frac{\partial F_j}{\partial \boldsymbol{\psi}_\Omega}\cdot\left(\frac{\partial \mathbf{H}_\Omega}{\partial \boldsymbol{\psi}_\Omega}\right)^{-1}}_{=\left[\mathbf{a}_{\Omega_j}^T, \text{c.c.}\right]}\cdot\frac{\partial \mathbf{H}_\Omega}{\partial s_n} = -\chi_e\frac{\partial}{\partial s_n}\left\{\mathbf{e}^\dagger\left[\mathbf{I}_3\otimes\text{diag}\left(\partial_j\mathbf{s}\right)\right]\mathbf{e}\right\} - 2\mathcal{R}\left[\mathbf{a}_{\Omega_j}^T\cdot\frac{\partial}{\partial s_n}\left(\hat{\mathbf{h}}[\mathbf{s}]\mathbf{e}\right)\right] =$$

$$\chi_e\partial_{j,trans}|\mathbf{e}|_n^2 + 2k^2\chi_e\mathcal{R}\left(\mathbf{a}_{\Omega_{j,trans},n}^T\cdot\mathbf{e}_n\right), \qquad \text{(S22)}$$

$$\frac{\delta T_j}{\delta s_n} = \chi_e\partial_{j,rot}|\mathbf{e}|_n^2 + 2k^2\chi_e\mathcal{R}\left(\mathbf{a}_{\Omega_{j,rot},n}^T\cdot\mathbf{e}_n\right), \qquad \text{(S23)}$$

where the partial gradient terms correspond to the following intensity gradients:

$$\partial_{j,trans}|\mathbf{e}|^2 \equiv \lim_{\delta x \to 0} \frac{|\mathbf{E}(\mathbf{r}-\delta x\hat{\jmath})|^2 - |\mathbf{E}(\mathbf{r}+\delta x\hat{\jmath})|^2}{2\delta x}, \tag{S24}$$

$$\partial_{j,rot}|\mathbf{e}|^2 \equiv \lim_{\delta\theta \to 0} \frac{|\mathbf{E}(\mathbf{R}(-\delta\theta\hat{\jmath})\cdot\mathbf{r})|^2 - |\mathbf{E}(\mathbf{R}(\delta\theta\hat{\jmath})\cdot\mathbf{r})|^2}{2\delta\theta}. \tag{S25}$$

To determine the adjoint field, we compute $\partial F_j/\partial\boldsymbol{\psi}_\Omega = [\partial F_j/\partial\mathbf{e}, \text{c.c.}]$ and $\partial\mathbf{H}_\Omega/\partial\boldsymbol{\psi}_\Omega$. First, note that $\partial F_j/\partial\mathbf{e} = -\chi_e\mathbf{e}^\dagger[\mathbf{I}_3 \otimes \text{diag}(\partial_j\mathbf{s})]$. Second, $\partial\mathbf{H}_\Omega/\partial\boldsymbol{\psi}_\Omega$ satisfies the following relation[9,10]:

$$\frac{\partial\mathbf{H}_\Omega}{\partial\boldsymbol{\psi}_\Omega} = \begin{bmatrix} \partial(\hat{\mathbf{h}}[\mathbf{s}]\mathbf{e})/\partial\mathbf{e} & \partial(\hat{\mathbf{h}}[\mathbf{s}]\mathbf{e})/\partial\mathbf{e}^* \\ \partial(\hat{\mathbf{h}}[\mathbf{s}]\mathbf{e})^*/\partial\mathbf{e} & \partial(\hat{\mathbf{h}}[\mathbf{s}]\mathbf{e})^*/\partial\mathbf{e}^* \end{bmatrix} = \begin{bmatrix} \hat{\mathbf{h}}[\mathbf{s}] & \mathbf{0} \\ \mathbf{0} & \hat{\mathbf{h}}[\mathbf{s}] \end{bmatrix}. \tag{S26}$$

The adjoint field $\mathbf{a}_{\Omega j}$ then satisfies the following equation:

$$\hat{\mathbf{h}}[\mathbf{s}]^T\mathbf{a}_{\Omega_j} = \nabla \times [\nabla \times \mathbf{a}_{\Omega_j}] - k^2[\mathbf{I}_3 \otimes \text{diag}(1+\chi_e\mathbf{s})]\mathbf{a}_{\Omega_j} = \left(\frac{\partial F_j}{\partial\mathbf{e}}\right)^T = -\chi_e[\mathbf{I}_3 \otimes \text{diag}(\partial_j\mathbf{s})]\mathbf{e}^*. \tag{S27}$$

Its solution can be analytically derived from the derivative of Eq. (S11) with respect to the infinitesimal translational and rotational displacement of the particle:

$$\partial_j'(\hat{\mathbf{h}}[\mathbf{s}]\mathbf{e}) = \nabla \times [\nabla \times \partial_j'\mathbf{e}] - k^2[\mathbf{I}_3 \otimes \text{diag}(1+\chi_e\mathbf{s})]\partial_j'\mathbf{e} + k^2\chi_e[\mathbf{I}_3 \otimes \text{diag}(\partial_j\mathbf{s})]\mathbf{e} = \mathbf{0}. \tag{S28}$$

Comparing Eq. (S28) to Eq. (S27) suggests that $\mathbf{a}_{\Omega_j} = \partial_j'\mathbf{e}^*/k^2$. A similar relation can be derived for the rotational modes.

### C. Gradient of detection efficiency

he gradient of $\eta_j[\partial_j'\mathbf{e}, \partial_j'\mathbf{e}^*]$ depends only on the functional derivatives of the electric fields, $\boldsymbol{\psi}_{\eta_j} = [(\partial_j'\mathbf{e})^T, \text{c.c.}]^T$. Similar to the approaches used in Eqs. (S20-23), we use $\mathbf{H}_{\eta_j} = [\partial_j'(\hat{\mathbf{h}}[\mathbf{s}]\mathbf{e})^T, \text{c.c.}]^T = \mathbf{0}$ as the corresponding Lagrangian functional to derive the following relation:

$$\frac{\delta\eta_j}{\delta s_n} = -\underbrace{\frac{\partial F_j}{\partial\boldsymbol{\psi}_{\eta_j}} \cdot \left(\frac{\partial\mathbf{H}_{\eta_j}}{\partial\boldsymbol{\psi}_{\eta_j}}\right)^{-1}}_{=[\mathbf{a}_{\eta_j}^T, \text{c.c.}]} \cdot \frac{\partial\mathbf{H}_{\eta_j}}{\partial s_n} = 2k^2\chi_e\mathcal{R}\left[\mathbf{a}_{\eta_j,n}^T \cdot (\partial_j'\mathbf{e} + \partial_j\mathbf{e})_n\right], \tag{S29}$$

where the adjoint field $\mathbf{a}_{\eta_j}$ satisfies the following inhomogeneous wave equation:

$$\left(\frac{\partial\left[\partial_j'(\hat{\mathbf{h}}[\mathbf{s}]\mathbf{e})\right]}{\partial\left(\partial_j'\mathbf{e}\right)}\right)\mathbf{a}_{\eta_j} = \nabla\times\left[\nabla\times\mathbf{a}_{\eta_j}\right] - k^2[\mathbf{I}_3\otimes\mathrm{diag}(1+\chi_e\mathbf{s})]\mathbf{a}_{\eta_j} = \left(\frac{\partial\eta_j}{\partial\left(\partial_j'\mathbf{e}\right)}\right)^T =$$

$$\left[\frac{\mathbf{P}_{det}}{\left(\partial_j'\mathbf{e}\right)^\dagger\mathbf{P}_{tot}\left(\partial_j'\mathbf{e}\right)} - \eta_j\mathbf{P}_{tot}\right]\left(\partial_j'\mathbf{e}^*\right) \quad \text{(S30)}$$

This equation was solved using the modified Born series. The discrete formulas of each optimization parameter and its gradient are summarized in Table S1.

| | **Discrete formula** |
|---|---|
| $m$ | $m[\mathbf{s}] = \rho_0\delta x^3\mathrm{tr}[\mathrm{diag}(\mathbf{s})]$,<br>$I_j[\mathbf{s}] = \rho_0\delta x^3\left(\lvert\mathbf{x}\rvert^2 - \mathbf{x}_j^2\right)^T\cdot\mathbf{s}$,<br>$\frac{\delta}{\delta s_n}[-(m-m_0)^2] - 2\rho_0\delta x^3(m-m_0)$. |
| $\Omega_j$ | $\Omega_{j,trans}^2 = \frac{F_j[\mathbf{s}\equiv S(\mathbf{r}+\delta x\hat{j})] - F_j[\mathbf{s}\equiv S(\mathbf{r}-\delta x\hat{j})]}{2m\delta x}$,<br>$F_j[\mathbf{s}\equiv S(\mathbf{r}-\delta\mathbf{r})] = -\chi_e\mathbf{e}^\dagger\left[\mathbf{I}_3\otimes\mathrm{diag}\left(\partial_j\mathbf{s}\right)\right]\mathbf{e}\big\rvert_{\mathbf{s}\equiv S(\mathbf{r}-\delta\mathbf{r})}$<br>$\Omega_{j,rot}^2 = \frac{T_j[\mathbf{s}\equiv S(\mathbf{R}(\delta\theta\hat{j})\cdot\mathbf{r})] - T_j[\mathbf{s}\equiv S(\mathbf{R}(-\delta\theta\hat{j})\cdot\mathbf{r})]}{2I_j\delta\theta}$,<br>$T_j[\mathbf{s}\equiv S(\mathbf{R}(-\delta\boldsymbol{\theta})\cdot\mathbf{r})] = -\chi_e\mathbf{e}^\dagger\left\{\mathbf{I}_3\otimes\mathrm{diag}\left[(\mathbf{x}\times\nabla\mathbf{s})_j\right]\right\}\mathbf{e}\big\rvert_{\mathbf{s}\equiv S(\mathbf{R}(-\delta\boldsymbol{\theta})\cdot\mathbf{r})}$<br>$\frac{\delta\Omega_{j,trans}^2}{\delta s_n} = -\frac{\rho_0\delta x^3\Omega_{j,trans}^2}{m} + \frac{1}{2m\delta x}\left[\frac{\delta F_j}{\delta s_n}\Big\rvert_{\mathbf{s}\equiv S(\mathbf{r}+\delta x\hat{j})} - \frac{\delta F_j}{\delta s_n}\Big\rvert_{\mathbf{s}\equiv S(\mathbf{r}-\delta x\hat{j})}\right]$,<br>$\frac{\delta\Omega_{j,rot}^2}{\delta s_n} = -\frac{\rho_0\delta x^3\Omega_{j,rot}^2\left(\lvert\mathbf{x}\rvert^2-\mathbf{x}_j^2\right)_n}{I_j} + \frac{1}{2I_j\delta\theta}\left[\frac{\delta T_j}{\delta s_n}\Big\rvert_{\mathbf{s}\equiv S(\mathbf{R}(\delta\theta\hat{j})\cdot\mathbf{r})} - \frac{\delta T_j}{\delta s_n}\Big\rvert_{\mathbf{s}\equiv S(\mathbf{R}(-\delta\theta\hat{j})\cdot\mathbf{r})}\right]$,<br>$\frac{\delta F_j}{\delta s_n} = \chi_e\partial_{j,trans}\lvert\mathbf{e}\rvert_n^2 + \chi_e\partial_{j,trans}'\lvert\mathbf{e}\rvert_n^2$, $\frac{\delta T_j}{\delta s_n} = \chi_e\partial_{j,rot}\lvert\mathbf{e}\rvert_n^2 + \chi_e\partial_{j,rot}'\lvert\mathbf{e}\rvert_n^2$. |
| $\eta_j$ | $\eta_j\left[\partial_j'\mathbf{e},\partial_j'\mathbf{e}^*\right] = \frac{\left(\partial_j'\mathbf{e}\right)^\dagger\mathbf{P}_{det}\left(\partial_j'\mathbf{e}\right)}{\left(\partial_j'\mathbf{e}\right)^\dagger\mathbf{P}_{tot}\left(\partial_j'\mathbf{e}\right)}$,<br>$\mathbf{P}_{tot} = \mathbf{U}^\dagger\{\mathbf{I}_3\otimes\mathrm{diag}[\delta(\lvert\mathbf{q}\rvert-k)]\}\mathbf{U}$,<br>$\mathbf{P}_{det} = \mathbf{U}^\dagger\{\mathbf{I}_3\otimes\mathrm{diag}[\delta(\lvert\mathbf{q}\rvert-k)\Theta(k\cdot\mathrm{NA}-\lvert\mathbf{q}_\perp\rvert)]\}\mathbf{U}$,<br>$\frac{\delta\eta_j}{\delta s_n} = 2k^2\chi_e\mathcal{R}\left[\mathbf{a}_{\eta_j,n}^T\cdot\left(\partial_j'\mathbf{e}+\partial_j\mathbf{e}\right)_n\right]$,<br>$\nabla\times\left[\nabla\times\mathbf{a}_{\eta_j}\right] - k^2[\mathbf{I}_3\otimes\mathrm{diag}(1+\chi_e\mathbf{s})]\mathbf{a}_{\eta_j} = \left[\frac{\mathbf{P}_{det}}{\left(\partial_j'\mathbf{e}\right)^\dagger\mathbf{P}_{tot}\left(\partial_j'\mathbf{e}\right)} - \eta_j\mathbf{P}_{tot}\right]\left(\partial_j'\mathbf{e}^*\right)$. |

**Table S1. Discrete forms of objective functionals and their gradients.**

**Supplementary Note 4. Shape constraint**

To limit the design freedom to simpler shapes, we introduce an interpolation function similar to the previous work[11], $\xi^{(n)}(\mathbf{r}) \equiv \boldsymbol{\xi}^{(n)} \in \mathbb{R}^{N\times 1}$, which is mapped onto the shape function by applying the sequential shape constraint operations. For 3D structure optimization, the following constraints were used:

$$\begin{aligned}
&\mathbf{s} = \max\left(\min\left(\boldsymbol{\xi}^{(3)}, 1\right), 0\right), && \text{(range limitation)}\\
&\boldsymbol{\xi}^{(3)} = \frac{1}{2}\left[1 + \frac{\tanh\left(\beta\left(\boldsymbol{\xi}^{(2)} - 1/2\right)\right)}{\tanh(\beta/2)}\right], && \text{(adaptive binarization)}\\
&\xi_q^{(2)} = \textstyle\sum_p D_{pq}\xi_p^{(1)}, && \text{(symmetry over XY, YZ, XZ planes)}\\
&\xi_p^{(1)} = \textstyle\sum_l C_{lp}\xi_l^{(0)}. && \text{(smoothing)}
\end{aligned} \tag{S31}$$

Here, $D_{pq}$ represents the linear mirror symmetry operator along XY, YZ, and XZ planes. $C_{lp}$ represents a Gaussian smoothing operator. The gradient of the objective functional with respect to the initial mapping function is by the chain rule $\delta\mathcal{L}/\delta\xi_l^{(0)} = \sum_n \left(ds_n/d\xi_l^{(0)}\right)\left(\partial\mathcal{L}/\delta s_n\right)$, where:

$$\frac{ds_n}{d\xi_l^{(0)}} = \sum_{p,q} \frac{\partial\xi_p^{(1)}}{\partial\xi_l^{(0)}}\frac{\partial\xi_q^{(2)}}{\partial\xi_p^{(1)}}\frac{\partial\xi_q^{(3)}}{\partial\xi_q^{(2)}}\frac{\partial s_n}{\partial\xi_q^{(3)}} = \frac{\beta\left[1-4\tanh^2(\beta/2)\right]}{2\tanh(\beta/2)}\left\{\sum_{p,q} C_{lp}D_{pq}\left(s_q - \frac{1}{2}\right)^2 \Theta\left[s_q\left(1-s_q\right)\right]\delta_{qn}\right\}. \tag{S32}$$

For extruded structures with the thickness of $y_0$, the following shape constraints were used:

$$\begin{aligned}
&\mathbf{s} = \Theta(y_0/2 - |\mathbf{y}|)\boldsymbol{\xi}_{\mathrm{XZ}}^{(f)}, && \text{(2D to 3D mapping)}\\
&\boldsymbol{\xi}_{\mathrm{XZ}}^{(f)} = \max\left(\min\left(\boldsymbol{\xi}_{\mathrm{XZ}}^{(3)}, 1\right), 0\right), && \text{(range limitation)}\\
&\boldsymbol{\xi}_{\mathrm{XZ}}^{(3)} = \frac{1}{2}\left[1 + \frac{\tanh\left(\beta\left(\boldsymbol{\xi}_{\mathrm{XZ}}^{(2)} - 1/2\right)\right)}{\tanh(\beta/2)}\right], && \text{(adaptive binarization)}\\
&\xi_{\mathrm{XZ},q}^{(2)} = \textstyle\sum_p D_{pq}\xi_{\mathrm{XZ},p}^{(1)}, && \text{(symmetry over XY, YZ, XZ planes)}\\
&\xi_{\mathrm{XZ},p}^{(1)} = \textstyle\sum_l C_{lp}\xi_{\mathrm{XZ},l}^{(0)}, && \text{(smoothing)}
\end{aligned} \tag{S33}$$

where $\boldsymbol{\xi}_{\mathrm{XZ}}^{(n)} \equiv B^{(n)}(x,z)$ denotes a 2D interpolation function that defines the base profile of an extruded object in the XZ plane. The gradient of the objective function satisfies $\delta\mathcal{L}/\delta\xi_{\mathrm{XZ},l}^{(0)} = \sum_n \left(ds_n/d\xi_{\mathrm{XZ},l}^{(0)}\right)\left(\partial\mathcal{L}/\delta s_n\right)$, where:

$$\frac{ds_n}{d\xi_{\mathrm{XZ},l}^{(0)}} = \frac{\beta y_0[1-4\tanh^2(\beta/2)]}{2\tanh(\beta/2)}\left\{\sum_{p,q} C_{lp}D_{pq}\left(\xi_{\mathrm{XZ},q}^{(f)}-\frac{1}{2}\right)^2 \Theta\left[\xi_{\mathrm{XZ},q}^{(f)}\left(1-\xi_{\mathrm{XZ},q}^{(f)}\right)\right]\delta_{qn}\right\}. \tag{S34}$$

## S5. Code implementation

Based on the gradient update rules mentioned above, we implement our algorithm. The pseudo-code for optimization with extruded shape constraints is shown in Table S2. The used parameters and the obtained results are available online.

```
% Fixed variables
λ = 1.55 μm; NA = 0.8; P = 250 mW; χ_e = 1.1 (SiO2) | 11.1 (Si);
Δx = 3D field of view; δx = voxel size = 50 nm;
P_det, P_tot = far-field projectors over 40 × 40 grids over steradian
z0 = thickness of the particle

% Initialise 3D field & particle susceptibility distribution
ξ_{XZ,j}^(0) = initialize_susceptibility(initial bead radius r0, standard deviation of background σ0);
e_in = initialize_standing_wave_trap(λ, NA, P, Δx, δx);

% Optimisation parameters
α = step size; τ_R = weight for rotation angular frequency;
τ_{m,1} = weight for m; τ_{m,2} = Additional mass weight;
s_th = threshold constant for τ_{m,2}

C_lp = (1 − δ_{jk,0}) / sqrt(π σ_s^2) · exp(−r_jk^2 / σ_s^2) + δ_{jk,0} = Smoothing filter with smoothing radius σ_s.
γ(Δp) = Increment factor of β per Δp-th iterations for adaptive binarization;

iter. = 1; % Begin optimisation
while iter. ≤ iter_max:
    % Shape constraint
    ξ_{XZ,p}^(f) = Σ_l C_lp ξ_{XZ,l}^(0);  ξ_{XZ,q}^(f) ← Σ_p D_pq ξ_{XZ,p}^(f);
    β = γ^⌊p/Δp⌋;  ξ_XZ^(f) ← 1/2 [1 + tanh(β(ξ_XZ^(f) − 1/2)) / tanh(β/2)];
    ξ_XZ^(f) ← max(min(ξ_XZ^(f), 1), 0);
    s = Θ(y0/2 − |y|) ξ_XZ^(f);

    % Forward solver – Omit detection efficiency computation for convenience
    e = solve_eq_S11(e_in, s); % Compute total electric field when particle.

    % Discrete objective functionals
      m[s] = ρ0 δx^3 tr[diag(s)];   I_j[s] = ρ0 δx^3 (|x|^2 − x_j^2)^T · s;
      F_j[s ≡ S(r − δr)] = −χ_e e† [I_3 ⊗ diag(∂_j s)] e |_{s≡S(r−δr)};
```

$$T_j[\mathbf{s} \equiv S(\mathbf{R}(-\delta\boldsymbol{\theta}) \cdot \mathbf{r})] = -\chi_e \mathbf{e}^\dagger \{\mathbf{I}_3 \otimes \mathrm{diag}[(\mathbf{x} \times \nabla\mathbf{s})_j]\}\mathbf{e}\big|_{\mathbf{s} \equiv S(\mathbf{R}(-\delta\boldsymbol{\theta})\cdot\mathbf{r})};$$

$$\Omega_{j,trans}^2 = \{F_j[\mathbf{s} \equiv S(\mathbf{r} + \delta x\hat{j})] - F_j[\mathbf{s} \equiv S(\mathbf{r} - \delta x\hat{j})]\}/(2m\delta x);$$

$$\Omega_{j,rot}^2 = \{T_j[\mathbf{s} \equiv S(\mathbf{R}(\delta\theta\hat{j}) \cdot \mathbf{r})] - T_j[\mathbf{s} \equiv S(\mathbf{R}(-\delta\theta\hat{j}) \cdot \mathbf{r})]\}/(2I_j\delta\theta);$$

% Inverse solver – Omit detection efficiency gradient for simplicity

$$\frac{\delta\mathcal{L}_m}{\delta s_n} = -2\rho_0\delta x^3(m - m_0)[\tau_{m,1} + \tau_{m,2}(s_n \geq s_{th})];$$ % add $\tau_{m,2}$ to fix high-RI region

$$\frac{\delta\Omega_{j,trans}^2}{\delta s_n} = -\frac{\rho_0\delta x^3\Omega_{j,trans}^2}{m} + \frac{1}{2m\delta x}\left[\frac{\delta F_j}{\delta s_n}\bigg|_{\mathbf{s}\equiv S(\mathbf{r}+\delta x\hat{j})} - \frac{\delta F_j}{\delta s_n}\bigg|_{\mathbf{s}\equiv S(\mathbf{r}-\delta x\hat{j})}\right];$$

$$\frac{\delta F_j}{\delta s_n} = \chi_e\partial_{j,trans}|\mathbf{e}|_n^2 + \chi_e\partial_{j,trans}'|\mathbf{e}|_n^2;$$

$$\partial_{j,trans}|\mathbf{e}|^2 = \{|\mathbf{e} \equiv \mathbf{E}(\mathbf{r} - \delta x\hat{j})|^2 - |\mathbf{e} \equiv \mathbf{E}(\mathbf{r} + \delta x\hat{j})|^2\}/(2\delta x);$$

$$\partial_{j,trans}'|\mathbf{e}|^2 = \{|\mathbf{e}[\mathbf{s} \equiv S(\mathbf{r} - \delta x\hat{j})]|^2 - |\mathbf{e}[\mathbf{s} \equiv S(\mathbf{r} + \delta x\hat{j})]|^2\}/(2\delta x);$$

$$\frac{\delta\Omega_{j,rot}^2}{\delta s_n} = -\frac{\rho_0\delta x^3\Omega_{j,rot}^2\left(|\mathbf{x}|^2 - \mathbf{x}_j^2\right)_n}{I_j} + \frac{1}{2I_j\delta\theta}\left[\frac{\delta T_j}{\delta s_n}\bigg|_{\mathbf{s}\equiv S(\mathbf{R}(\delta\theta\hat{j})\cdot\mathbf{r})} - \frac{\delta T_j}{\delta s_n}\bigg|_{\mathbf{s}\equiv S(\mathbf{R}(-\delta\theta\hat{j})\cdot\mathbf{r})}\right];$$

$$\frac{\delta T_j}{\delta s_n} = \chi_e\partial_{j,rot}|\mathbf{e}|_n^2 + \chi_e\partial_{j,rot}'|\mathbf{e}|_n^2;$$

$$\partial_{j,rot}|\mathbf{e}|^2 = \{|\mathbf{e} \equiv \mathbf{E}(\mathbf{R}(-\delta\theta\hat{j}) \cdot \mathbf{r})|^2 - |\mathbf{e} \equiv \mathbf{E}(\mathbf{R}(\delta\theta\hat{j}) \cdot \mathbf{r})|^2\}/(2\delta\theta);$$

$$\partial_{j,rot}'|\mathbf{e}|^2 = \{|\mathbf{e}[\mathbf{s} \equiv S(\mathbf{R}(-\delta\theta\hat{j}) \cdot \mathbf{r})]|^2 - |\mathbf{e}[\mathbf{s} \equiv S(\mathbf{R}(\delta\theta\hat{j}) \cdot \mathbf{r})]|^2\}/(2\delta\theta);$$

$$\frac{\delta\mathcal{L}_\Omega}{\delta s_n} = \sum_j\left[\frac{\delta\Omega_{j,trans}^2}{\delta s_n} + \tau_R\frac{\delta\Omega_{j,rot}^2}{\delta s_n}\right];$$ % Different weight to rotation modes

% Shape update

$$\frac{\delta\mathcal{L}}{\delta\xi_{\mathrm{XZ},l}^{(0)}} = \sum_n\frac{ds_n}{d\xi_{\mathrm{XZ},l}^{(0)}}\left(\frac{\delta\mathcal{L}_\Omega}{\delta s_n} + \frac{\delta\mathcal{L}_m}{\delta s_n}\right);\ \boldsymbol{\xi}_{\mathrm{XZ}}^{(0)} \leftarrow \boldsymbol{\xi}_{\mathrm{XZ}}^{(0)} + \frac{\alpha}{\chi_e}\frac{\delta\mathcal{L}}{\delta\xi_{\mathrm{XZ},l}^{(0)}};$$

*iter.* ← *iter.* + 1;

**end**

% Termination with binarization & parameter estimations

$\mathbf{s} \leftarrow \mathbf{s} \geq 0.5$; Get $\mathbf{e}$, $\Omega[\mathbf{e}, \mathbf{e}^*, \mathbf{s}]$, $m[\mathbf{s}]$, $I_j[\mathbf{s}]$, and $\eta_j[\partial_j'\mathbf{e}, \partial_j'\mathbf{e}^*]$;

**Table S2.** Pseudo-code for optimizing a 3D extruded object.